\documentclass{revtex4}

\usepackage{graphicx}

\begin{document}


\title{Analytical Wake Potentials in a Closed Pillbox Cavity}

\thanks{
This work was supported by the U.S. Department of Energy grant
DE-FG02-04ER41317.
}

\author{Gregory~R.~Werner}
\email{Greg.Werner@colorado.edu}

\affiliation{Center for Integrated Plasma Studies, University of
Colorado, Boulder, Colorado 80309}

\begin{abstract}

Wake potentials are derived for a closed (cylindrical) pillbox
cavity as a sum over cavity modes.  The resulting expression applies
to on- and off-axis beams and test particles.
The sum is evaluated numerically for
a Gaussian drive bunch and compared to the wake potential derived from
simulation.

\end{abstract}

\maketitle


\section{Introduction}

The wake potential \cite{BaneWakeFieldGuide,WilsonWakeFieldGuide}
describes the interaction between two charged
particles mitigated by an external structure; the wake
potential is especially useful for considering
charged particles traveling at relativistic speeds on parallel
paths through structures (such as radio-frequency accelerating cavities)
in a particle accelerator.  In this paper we derive the 
longitudinal wake potential for highly relativistic beams traveling
parallel to the axis---but not necessarily on the axis---of 
a closed (cylindrical) pillbox cavity.

The monopole wake potential for on-axis beams in a closed pillbox cavity
was derived in \cite{Weiland:1981} (see also 
\cite{BaneWakeFieldGuide,WilsonWakeFieldGuide}).
Analytical wake potentials have also been found for other geometries,
such as closed spherical cavities \cite{Ratschow:2002},
conical cavities \cite{Tsakanian:2005},
multi-cell pillbox cavities with elliptical cross section (for
axial beams only) \cite{Kim:1990},
and multi-cell pillbox cavities with beam tubes
\cite{Gao:1995,Gao:1996}.
We note that the wake potential for a pillbox cavity with 
infinitely long beam tubes differs significantly from the
wake potential in a closed pillbox cavity: in a pillbox cavity
with infinitely long beam tubes, the 
parts of the wake potential that vary azimuthally as $\cos(m\theta)$
vary radially as $r^m$;
in a closed pillbox cavity, the radial dependence is more complicated.

\section{Wake potential}

The wake potential describes the momentum change of a test particle
caused by the fields excited by another charged particle.
When a point charge $q_b$ ($b$ for ``beam'') 
travels through a cavity, it creates an electric field
$\mathbf{E}(r,\theta,z,t)$ and also a magnetic field;
the wake potential describes the momentum change of
a test charge due to those fields.

The wake potential
is particularly useful when applied to highly relativistic particles, 
which maintain nearly constant speed while changing momentum.
If a charge $q_b$ travels at light speed along the path
$(r,\theta,z)(t) = (r_b,\theta_b, ct)$, creating an
electric field with $z$-component $E_z(r,\theta,z,t)$, its
longitudinal wake potential is:
\begin{equation}
   W_z(s;r_b,\theta_b, r_t,\theta_t) 
   \equiv -\frac{1}{q_b} \int dz\, E_z(r_t, \theta_t, z, t\!=\!(z+s)/c)
.\end{equation}
(We will consider a closed cavity of length $\ell$, so the integration
runs from $z=0$ to $z=\ell$.)
If a test charge $q_t$ trails $q_b$ by a distance $s$, 
traveling along a parallel path
$(r,\theta,z)(t) = (r_t,\theta_t, ct+s)$ from $z=0$ to $z=\ell$, 
its longitudinal
momentum change due to the fields created by $q_b$ is
\begin{equation}
   \Delta p_z(s;r_t,\theta_t) = -\frac{q_t q_b}{c}
     W_z(s;r_b,\theta_b,r_t,\theta_t)
\end{equation}
(the test charge is assumed to be highly relativistic, so its
speed remains constant even as its momentum changes).
By convention, positive $W_z$ corresponds to a loss in momentum.

In cylindrically symmetric structures, it is helpful to decompose wakefields into
azimuthal harmonics, i.e., fields with dependence $\cos(m\theta)$
and $\sin(m\theta)$ for $m=0,1,\ldots$:
\begin{equation}
	W_z(s;r_b,r_t,\theta_t) = \sum_{m=0}^\infty W_{z,m}(s;r_b,r_t)
	  \cos(m \theta_t)
\end{equation}
where we have chosen the beam
to be at $\theta_b=0$; symmetry prohibits the excitation of
any modes with $\sin(m\theta)$ dependence.
In structures with cylindrical symmetry \emph{and}
infinitely extended beam tubes, the azimuthal components
of the wakefield have a particularly simple dependence on $r_b$
and $r_t$ (when they are smaller than the beam tube radius)
\cite{Weiland:1983}
\begin{equation}
	W_{z,m}(s;r_b,r_t) \propto r_b^m r_t^m 
.\end{equation}
However, Ref.~\cite{Napoly:1993} shows that this simple form depends
on the fields at $z=\pm \infty$ being the fields of a charge in an
infinitely long beam tube; in a closed cavity, the fields
at the ends of integration (at $z=0$ and $z=\ell$) 
upset the simple dependence on $r_b$ and $r_t$, though $W_{z,m}$ still
approximately assumes the above form for $r_b$ and $r_t$ near the axis.

\section{Pillbox Wake Potential}
\label{sec:AnalyticPillbox}

The monopole wake potential for a closed pillbox cavity
has been derived for axial beam and test particles 
in \cite{Weiland:1981}.
In the same way, we derive an
analytical expression for multipole wakes, for beam and
test particles following paths parallel to the axis.
We will write the wake
potential as a sum over cavity modes; only TM modes (with magnetic
field transverse to the $z$ direction) will be excited,
since the beam travels in the $z$
direction and TE modes have $E_z=0$.
Choosing the beam to be at $\theta_b=0$, only TM modes
with $\cos(m\theta)$ dependence will be excited, and we need not
consider modes with $\sin(m\theta)$ dependence.

We can write the field in the pillbox cavity of radius $R$ and
length $\ell$ as a
sum over modes; the (cosine) mode $\textrm{TM}_{mnp}$ 
(for integers $m \geq 0$, $n\geq 1$, $p\geq 0$) has fields 
(see \cite{AcceleratorHandbook}, Sec.~7.3.7):
\begin{eqnarray}
  E_{z,mnp}(r,\theta,z,t) &=& 
      \cos \left( \frac{p\pi z}{\ell} \right)
	J_m \left( \frac{j_{m,n} r}{R} \right) 
	\cos(m\theta)
	e^{-i\omega_{mnp}t}
	\\
  E_{r,mnp}(r,\theta,z,t) &=&
	-\left( \frac{p\pi}{\ell} \right) \frac{R}{j_{m,n}}
	\sin\left( \frac{p\pi z}{\ell} \right)
	     J'_m \left( \frac{j_{m,n} r}{R} \right) \cos(m\theta)
	e^{-i\omega_{mnp} t} 
	\\
  E_{\theta,mnp}(r,\theta,z,t) &=&
	\left( \frac{p\pi}{\ell} \right) \frac{R^2}{j_{m,n}^2}
	\sin\left( \frac{p\pi z}{\ell} \right)
	 \frac{m}{r} 
	     J_m \left( \frac{j_{m,n} r}{R} \right) \sin(m\theta)
	e^{-i\omega_{mnp} t} 
	\\
  H_{r,mnp}(r,\theta,z,t) &=& 
	i \sqrt{\frac{\epsilon_0}{\mu_0}}
	\frac{\omega_{mnp} R^2}{c j_{m,n}^2}
	\cos\left( \frac{p\pi z}{\ell} \right)
	\frac{m}{r} 
	     J_m \left( \frac{j_{m,n} r}{R} \right) \sin(m\theta)
	e^{-i\omega_{mnp} t} 
	\\
  H_{\theta,mnp}(r,\theta,z,t) &=& 
	i \sqrt{\frac{\epsilon_0}{\mu_0}}
	\frac{\omega_{mnp} R}{c j_{m,n}}
	\cos\left( \frac{p\pi z}{\ell} \right)
	     J'_m \left( \frac{j_{m,n} r}{R} \right) \cos(m\theta)
	e^{-i\omega_{mnp} t} 
\end{eqnarray}
where $J_m$ is the Bessel function of order $m$, $J_m'$ is its
derivative, and $j_{m,n}$ is the $n^\textrm{th}$ zero of $J_m$.
Mode $\textrm{TM}_{mnp}$ oscillates with frequency $\omega_{mnp}$:
\begin{equation}
  \frac{\omega_{mnp}^2}{c^2} = \frac{j_{m,n}^2}{R^2} + 
  \left( \frac{p\pi}{\ell} \right)^2 
.\end{equation}

The wake potential for a highly relativistic
point charge traveling parallel to the $z$
axis, at radius $r_b$ (and $\theta_b=0$), is \cite{WilsonWakeFieldGuide}:
\begin{equation} \label{eq:Wz}
  W_z(s;r_b,r_t,\theta_t) = 2 H(s) \sum_{mnp} k_{mnp}(r_b,r_t,\theta_t)
   \cos \left( \frac{\omega_{mnp} s}{c} \right)
\end{equation}
where $H(s)$ is the Heaviside step function, and $H(0)=1/2$,
and $k_{mnp}(r_b, r_t,\theta_t)$ 
is the loss factor for mode $\textrm{TM}_{mnp}$:
\begin{equation}
	k_{mnp}(r_b,r_t,\theta_t) = \frac{
	  V_{mnp}^* (r_b,\theta_b=0) V_{mnp}(r_t,\theta_t) }{ 4 U_{mnp} }
\end{equation}
where $V_{mnp}(r,\theta)$ is the (complex) voltage gain of a test particle
crossing the cavity at transverse position $(r,\theta)$, due to mode
$\textrm{TM}_{mnp}$ when the cavity has stored energy $U_{mnp}$ in that mode:
\begin{eqnarray}
  && V_{mnp}(r,\theta) = \int_0^\ell dz \, E(r,\theta,z,t=z/c)
    \\
	&&= 
	J_m \left( \frac{j_{m,n} r}{R} \right) 
	\cos(m\theta)
	\frac{i\omega_{mnp} R^2}{c j_{m,n}^2} 
       \left[ (-1)^p e^{-i\omega_{mnp} \ell/c} -1 \right]
	 \nonumber
\end{eqnarray}
and
\begin{eqnarray}
	U_{mnp} &=&
	\frac{1+\delta_{m0}}{2-\delta_{p0}}
	   \frac{\pi \epsilon_0 R^4 \ell}{4j_{m,n}^2} J'_m(j_{m,n})^2
	   \frac{\omega_{mnp}^2}{c^2} 
\end{eqnarray}
where $\delta_{mn}$ is the Kronecker delta.
The loss factor is then:
\begin{eqnarray}
	k_{mnp}(r_b, r_t, \theta_t) &=&
	 \frac{2-\delta_{p0}}{1+\delta_{m0}}
	\frac{
	 J_m \left( \frac{j_{m,n} r_b}{R} \right) 
	 J_m \left( \frac{j_{m,n} r_t}{R} \right) 
	 \cos(m\theta_t)
       \cdot 2\left[ 1 - (-1)^p \cos(\omega_{mnp} \ell/c) \right]
	}{ \displaystyle
	   \pi \epsilon_0  \ell j_{m,n}^2 J'_m(j_{m,n})^2
	}
.\end{eqnarray}

The sum in Eq.~(\ref{eq:Wz}) unfortunately does not converge:
for fixed $m$ and $n$, the terms oscillate with constant
amplitude as $p$ increases.   In case
the beam and test particles travel along the same line, the sum
can be analytically evaluated for $s$ small enough that
the walls at $r=R$ can have no effect
\cite{Weiland:1981,WilsonWakeFieldGuide}---it 
sums to a sequence of delta functions, which
offers some explanation for the sum's lack of convergence
(considering the representation of a delta function as a 
non-convergent Fourier series).

The sum's behavior can be improved by calculating the wake potential
due to a charged bunch with total charge $q_b$ and linear 
density profile $\lambda(s)$,
rather than a point charge.  The bunch wake potential is
\begin{equation} \label{eq:bunchDensity}
	V_z(s) = \int_0^\infty ds' \lambda(s-s') W_z(s')
.\end{equation}
We will consider a Gaussian bunch
\begin{equation}
	\lambda(s) = \frac{1}{\sqrt{2\pi} \sigma} \exp
	\left( - \frac{s^2}{2\sigma^2} \right)
\end{equation}
and 
replace the $\cos(\omega_{mnp}s/c)$ term in Eq.~(\ref{eq:Wz}), 
with the integral
\begin{eqnarray}
	&&\int_0^\infty ds' \, \frac{1}{\sqrt{2\pi} \sigma}
	\exp \left( - \frac{(s-s')^2}{2\sigma^2} \right)
	\cos(ks')
	\nonumber \\
	&& = \frac{1}{2} 
	\exp \left( - \frac{\sigma^2k^2}{2} \right)
	\textrm{Re} \left[ 
	  e^{iks} \textrm{erfc}\left( -\frac{s+i\sigma^2 k}{\sqrt{2}\sigma}
	    \right) \right]
,\end{eqnarray}
where $k=\omega_{mnp}/c$,
and $\textrm{erfc}$ is one minus the error function (of a complex
argument). 
The bunch wake potential is:
\begin{eqnarray} \label{eq:bunchPotential}
  V_z(s;r_b,r_t,\theta_t) &=&
    2 H(s) \sum_{m,n,p}
	 \frac{2-\delta_{p0}}{1+\delta_{m0}} 
	 \nonumber\\
	 &&\times
	\frac{
	 J_m \left( \frac{j_{m,n} r_b}{R} \right) 
	 J_m \left( \frac{j_{m,n} r_t}{R} \right) 
	 \cos(m\theta_t)
       \cdot 2\left[ 1 - (-1)^p \cos(\omega_{mnp} \ell/c) \right]
	}{ \displaystyle
	   \pi \epsilon_0  \ell j_{m,n}^2 J'_m(j_{m,n})^2
	}
	\nonumber \\
	&& \times
	\frac{1}{2} 
	\exp \left( - \frac{\sigma^2 \omega_{mnp}^2}{2c^2} \right)
	\textrm{Re} \left[ 
	  e^{i\omega_{mnp}s/c} \textrm{erfc}\left( -\frac{s+i\sigma^2 \omega_{mnp}/c}{\sqrt{2}\sigma}
	    \right) \right]
.\end{eqnarray}
If we wish to know
just the contribution from modes with $\cos(m\theta)$ dependence,
we sum only over modes with that $m$.

For $x\ll m$, $J_m(x) \approx (x/2)^m/m!$ \cite{Abramowitz},
so the multipole contributions to the
wake are proportional to $r_t^m r_b^m$ when $r_b^m$ and $r_t^m$ are
small.

\section{Numerical Pitfalls}
\label{sec:erfcEval}

The bunch potential in Eq.~(\ref{eq:bunchPotential}) can be evaluated
in a straightforward manner, except for the 
function $e^{-y^2+2ixy}\textrm{erfc}(-x-iy)$, where we have used the
abbreviations
$x=s/(\sqrt{2}\sigma)$ and
$y=\sigma \omega_{mnp}/(\sqrt{2}c)$.  

An easy and fast way to evaluate
$\textrm{erfc}$ uses the Faddeeva function 
$w(z)\equiv e^{-z^2}\textrm{erfc}(-iz)$ computed by the method of 
\cite{Weideman:1994}; i.e., 
\begin{equation}
	e^{-y^2 + 2ixy} \textrm{erfc}(-x-iy) 
	 = e^{-x^2} w(y-ix)
.\end{equation}
However, $w(y-ix)$ becomes enormous for $x^2 \gg y^2$, while
$e^{-x^2}w(y-ix)$ remains tractable; similarly, for large $y^2$,
$\textrm{erfc}(-x-iy)$ is enormous.  In these cases
we use the asymptotic expansion for $\textrm{erfc}$ 
(see \cite{Abramowitz}); because the expansion in \cite{Abramowitz}
for $\textrm{erfc}(z)$
is valid for $|\textrm{arg} z|<3\pi/4$, we have to apply the identity
$\textrm{erfc}(-x-iy)=2-\textrm{erfc}(x+iy)$ before using the expansion
when $x > |y|$.

Specifically, to prevent overflow with double precision arithmetic, 
we perform a different evaluation for the following cases:
if $x^2-y^2>500$ and $x>0$, we evaluate
\begin{equation}
	e^{-y^2 + 2ixy} \textrm{erfc}(-x-iy) 
	\approx 2e^{-y^2+2ixy} 
\end{equation}
and if $x^2-y^2>500$ and $x<0$, or if $y^2-x^2>500$, then we evaluate
\begin{equation}
	e^{-y^2 + 2ixy} \textrm{erfc}(-x-iy) 
	 \approx -\frac{e^{-x^2}}{\sqrt{\pi}(x+iy)}
	\left[ 1 + \sum_{m=1}^{6} (-1)^m 
	    \frac{1\cdot 3 \ldots (2m-1)}{[2(x^2-y^2+2ixy)]^m}
	    \right]
.\end{equation}
These approximations yield nearly full accuracy in double precision 
when the specified conditions are satisfied.

\section{Behavior for large $n$ and $p$}

In this section we show the behavior of terms in the sum 
in Eq.~(\ref{eq:bunchPotential}) for
large $n$ and $p$.
For large $p$ ($\omega_{mnp}/c \gg 1/\sigma$),  we must consider
two cases (the asymptotic limits are given in Sec.~\ref{sec:erfcEval}):
the non-oscillatory contributions that depend on $p$ behave as
\begin{eqnarray}
	&& \exp \left( - \frac{\sigma^2 \omega_{mnp}^2}{2c^2} \right)
	\textrm{Re} \left[ 
	  e^{i\omega_{mnp}s/c} \textrm{erfc}\left( -\frac{s+i\sigma^2 \omega_{mnp}/c}{\sqrt{2}\sigma}
	    \right) \right]
	    \nonumber \\
	    && \sim
	    \left\{ \begin{array}{c@{\quad}l} 
    2 e^{-\sigma^2 (\omega_{mnp}/c)^2/2}  \cos(\omega_{mnp}s/c) 
           & \textrm{ if } s/\sigma \gg \sigma\omega_{mnp}/c
	     \\
   -\sqrt{\frac{2}{\pi}}
    \frac{ s/\sigma  }{
          (s/\sigma)^2 + (\sigma \omega_{mnp}/c)^2 }
	    e^{-s^2/2\sigma^2}
           & \textrm{ if } |s/\sigma | \ll \sigma\omega_{mnp}/c
          \end{array} \right.
\end{eqnarray}
When $s \gg \sigma$, convergence is very fast (with truncation
error falling as the tail of a Gaussian), 
since the Gaussian charge
distribution has had time to cancel out high-frequency contributions.
When $s$ is comparable to or smaller than $\sigma$, the wake potential
``feels'' only part of the charge distribution, and higher frequencies
matter more; consequently, the terms in the sum eventually fall off as
$1/\omega_{mnp}^2$.

For fixed $m$, but large $n$ ($n \gg mR/r_b$, $n \gg mR/r_t$),
the envelope behavior (ignoring oscillatory contributions)
of terms in the bunch potential series that depend on $n$ is:
\begin{equation}
	\frac{
	 J_m \left( \frac{j_{m,n} r_b}{R} \right) 
	 J_m \left( \frac{j_{m,n} r_t}{R} \right) 
	}{ \displaystyle
	   j_{m,n}^2 J'_m(j_{m,n})^2
	}
	\sim 
	\frac{R}{(n+m/2)^2 \pi^2 \sqrt{r_b r_t}}
,\end{equation}
decreasing slowly as $\sim 1/n^2$.

\section{Wake potential simulation}

We compared the analytical wake potential against 
$W_{z,m}(s;r_b,r_t)$ calculated via simulation, using the electromagnetic 
particle-in-cell (PIC) simulation capability of 
\textsc{vorpal} \cite{Nieter:2004} in Cartesian coordinates.  We excited a
cavity using a current bunch at $r_b$ (and $\theta_b=0$),
with a Gaussian width $\sigma$ in the longitudinal direction $z$;
the current bunch traveled at a highly relativistic speed and
was (artificially) unaffected by the fields it generated.  At each time step,
we injected highly relativistic test particles (with charges too small
to affect the cavity fields) at radius $r_t$ and regularly-spaced
angles $\theta_t$, and recorded the momentum change of each test particle
after crossing the cavity.
We thus measured $\Delta p_z(s;r_t,\theta_t)$, which, after decomposition
into azimuthal harmonics, yielded $W_{z,m}(s;r_b,r_t)$.

With a radius $R=11.5\:$mm,
the cavity's $\textrm{TM}_{010}$ mode oscillated at
10 GHz;
the length was chosen to be one-half wavelength at
that frequency, $\ell=15.0\:$mm.  The Gaussian bunch 
[as in Eq.~(\ref{eq:bunchDensity})] had $\sigma = 2\ell/25$ (but
was very thin in cross-section).
Following the advice of Ref.~\cite{AcceleratorHandbook} (Sec.~3.2.3),
we chose the cell length $\Delta z < \sqrt{\sigma^3/\ell}$,
resulting in $\Delta z = 0.333\:$mm and
transverse cell sizes $\Delta x=\Delta y = 0.336\:$mm.

The curved metal boundaries of the cavity were simulated using the
Dey-Mittra algorithm \cite{DeyMittra:1999}, which requires (for
improved accuracy) a reduction in time step from the standard
Courant-Friedrichs-Lewy time step.

\begin{figure}[tp]
\centering
(a)\includegraphics*[trim = 0.2in 0in 0.1in 0.5in,width=3.1in]{%
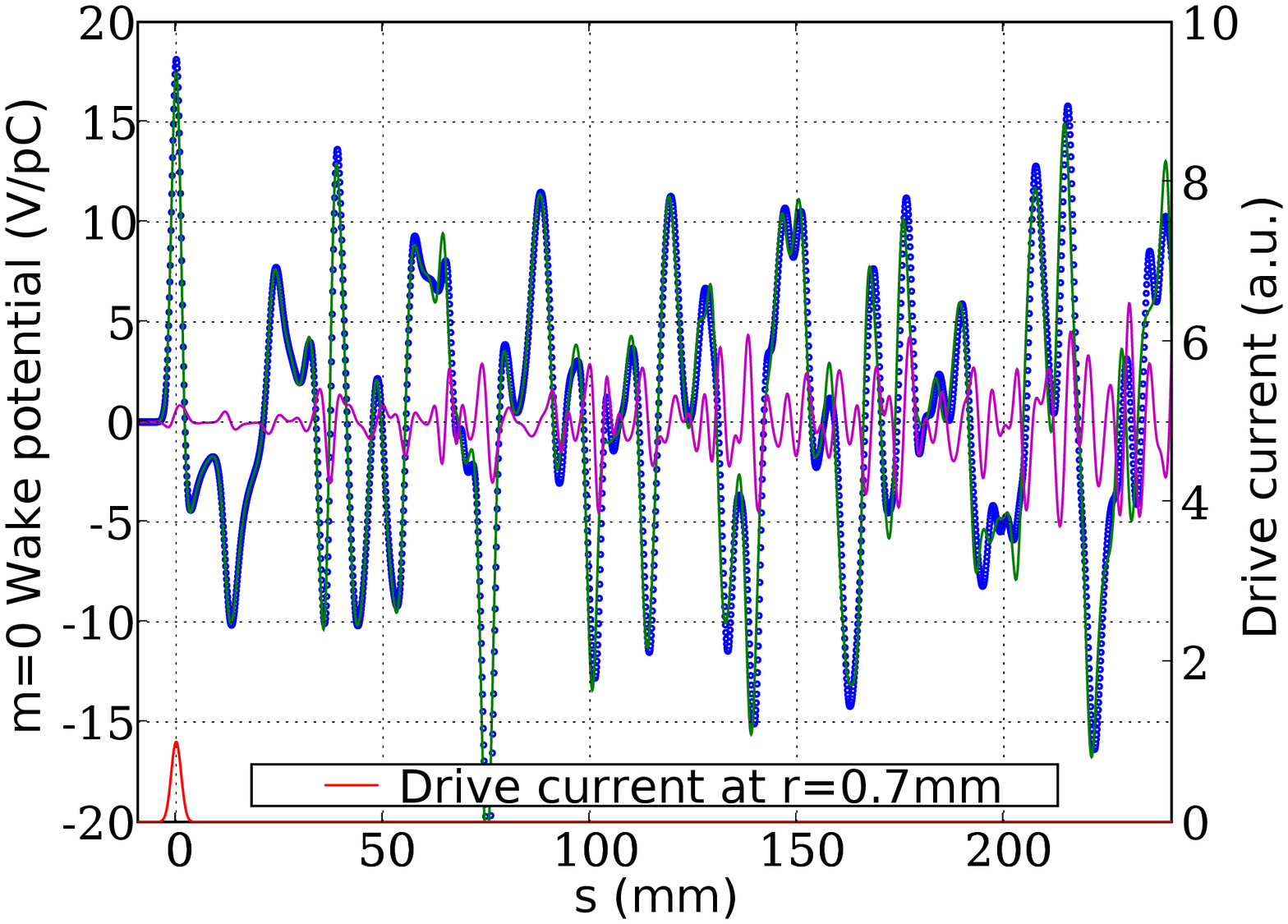}\\%
(b)\includegraphics*[trim = 0.1in 0in 0.1in 0.5in,width=3.1in]{%
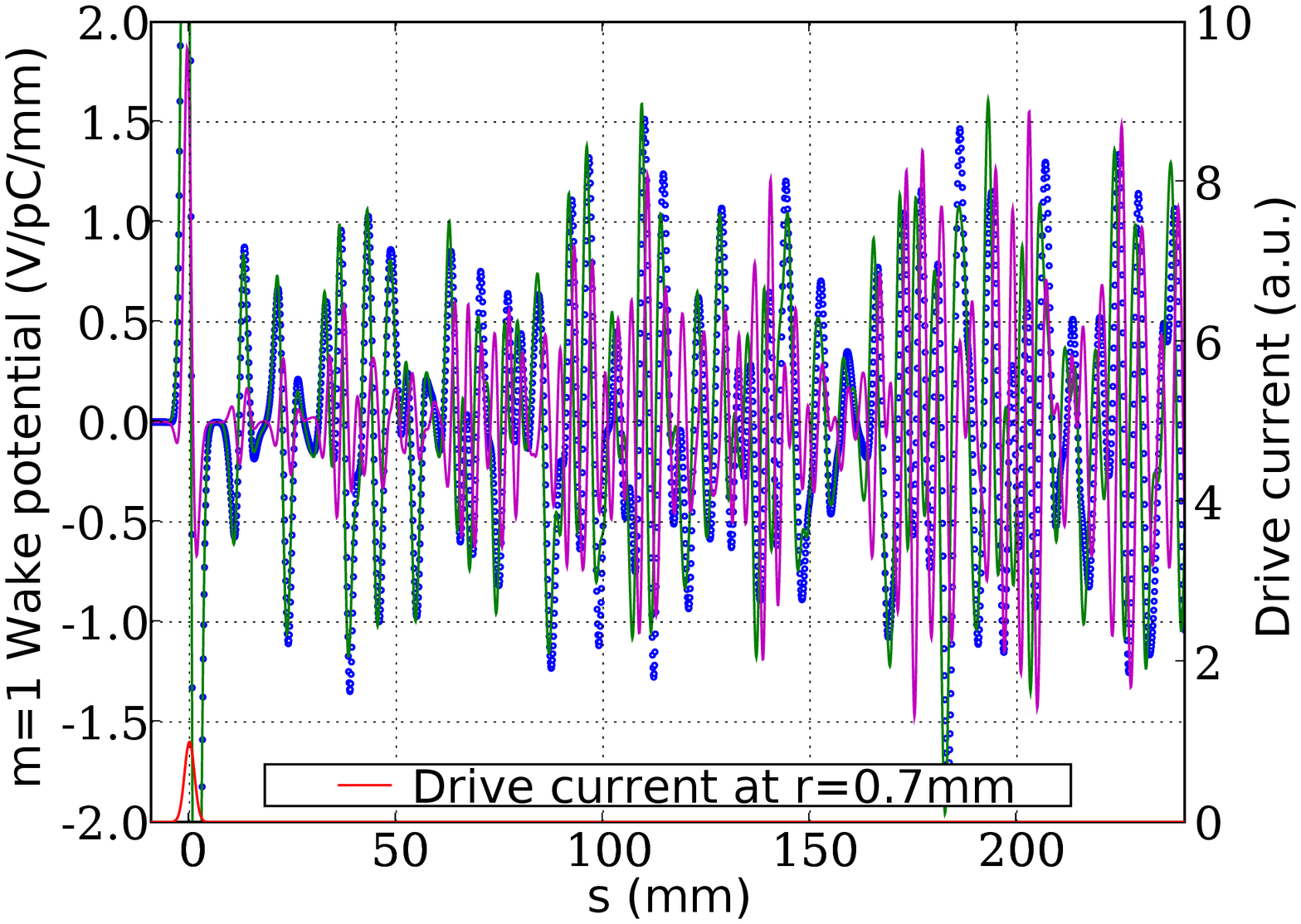}
\caption{Monopole wake potential (a) and dipole wake potential
divided by the test-particle radius (b) in the closed 
pillbox cavity: the analytical
(green line) and simulated (blue circles) wake potentials are nearly
the same; the difference between them is shown by the magenta line;
the red line at the bottom shows the drive current.
The analytical sum was cutoff above $\omega/c=300/\sigma$
($f\approx 1.2\times 10^{4}\:$GHz) for $s/\sigma<8.5$ and
$\omega/c=10/\sigma$ ($400\:$GHz) for larger $s$.
Simulated results are shown for cells of size $0.33\:$mm, and a
time step half the Courant-Friedrichs-Lewy time step.
\label{fig:wakes}}
\end{figure}

Figure~\ref{fig:wakes} shows the resulting $m=0$ and $m=1$ components
of the wake (bunch) potential for identical beam and test-particle radii, 
$r_b=r_t=0.672\:$mm.  The difference between the analytical and simulated
wakefield grows in time as the simulated modes slip in phase with
respect to the analytical modes (due to error in the frequency of the
simulated modes).

\begin{figure}[tp]
\centering
(a)\includegraphics*[trim = 0.2in 0in 0.1in 0.5in,width=3.1in]{%
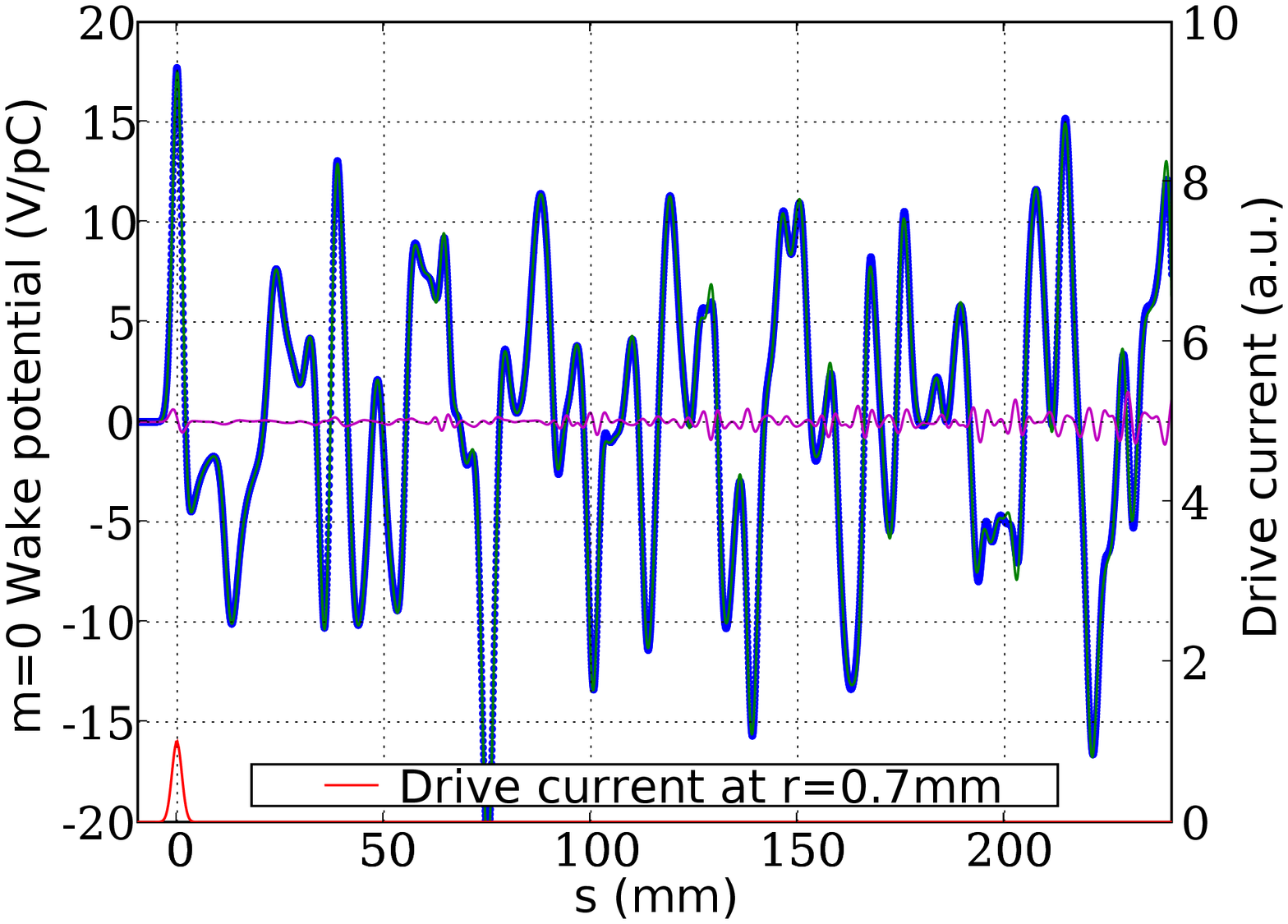}\\%
(b)\includegraphics*[trim = 0.1in 0in 0.1in 0.5in,width=3.1in]{%
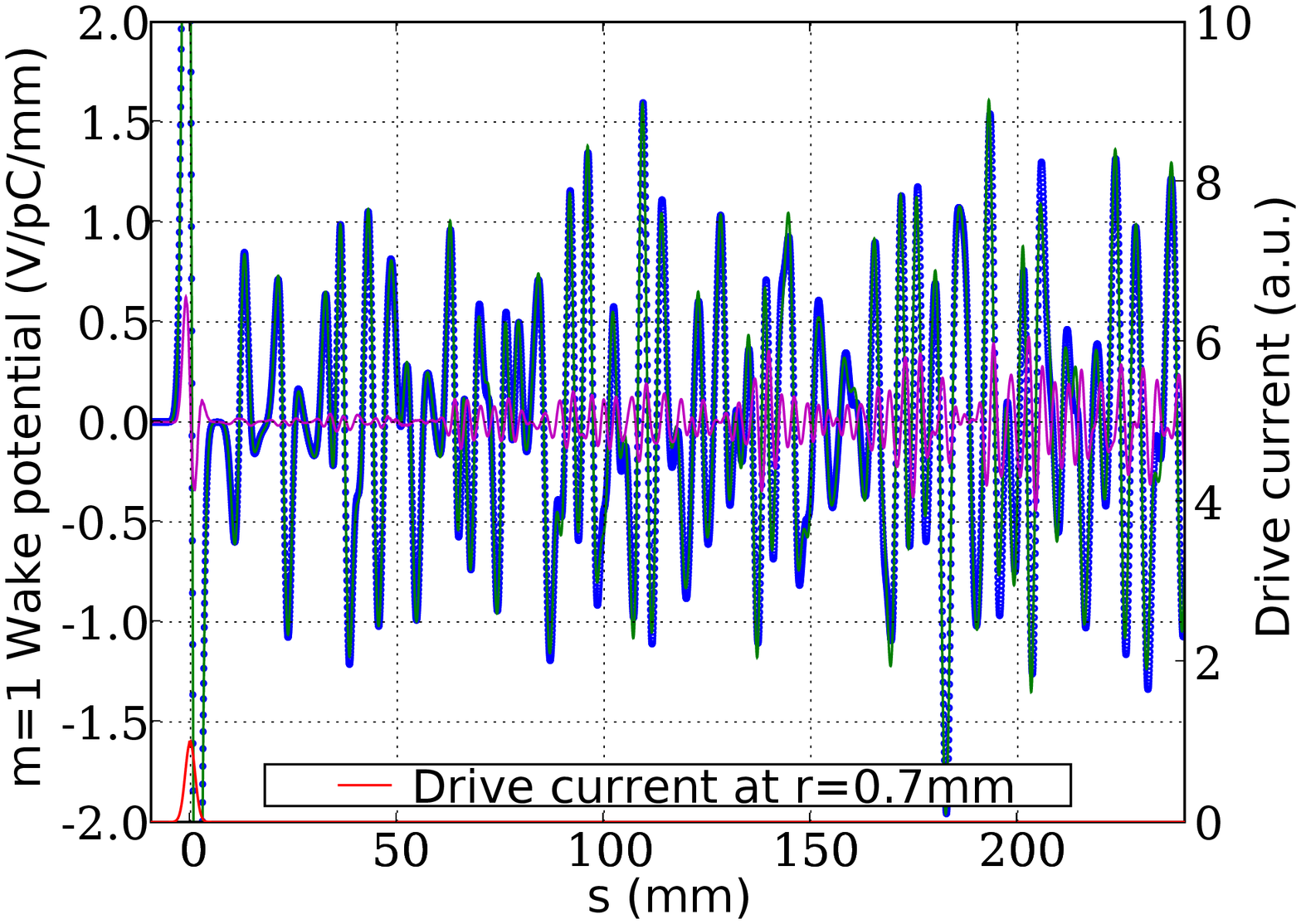}
\caption{Monopole wake potential (a) and dipole wake potential
divided by the test-particle radius (b) in the closed 
pillbox cavity: the analytical
(green line) and simulated (blue circles) wake potentials are nearly
the same; the difference between them is shown by the magenta line.
The analytical sum was cutoff above $\omega/c=300/\sigma$
($f\approx 1.2\times 10^{4}\:$GHz) for $s/\sigma<8.5$ and
$\omega/c=10/\sigma$ ($f\approx 400\:$GHz) for larger $s$.
Simulated results are shown for cells of size $0.17\:$mm, and a
time step 40\% of the Courant-Friedrichs-Lewy time step.
\label{fig:dblResWakes}}
\end{figure}

Figure~\ref{fig:dblResWakes} shows the wakefields compared to
a higher resolution simulation, with 
$\Delta_z \approx \Delta x=\Delta y = 0.17\:$mm.  The error between
simulation and theory is correspondingly reduced.

\begin{acknowledgments}

To compute wakefields we used the simulation framework \textsc{vorpal}, 
which was developed with support of
the Offices of HEP, FES, and NP of the Department of Energy, the
SciDAC program, AFOSR, JTO, Office of the Secretary of Defense, and
the SBIR programs of the Department of Energy and Department of
Defense.  
We would also like to acknowledge assistance from the rest of 
the \textsc{vorpal} team: 
T.~Austin, 
G.~I.~Bell, 
D.~L.~Bruhwiler, 
R.~S.~Busby, 
J.~Carlsson, 
J.~R.~Cary,
B.~M.~Cowan, 
D.~A.~Dimitrov, 
A.~Hakim, 
J.~Loverich, 
P.~Messmer, 
P.~J.~Mullowney, 
C.~Nieter, 
K.~Paul, 
S.~W.~Sides, 
N.~D.~Sizemore, 
D.~N.~Smithe, 
P.~H.~Stoltz, 
S.~A.~Veitzer, 
D.~J.~Wade-Stein, 
M.~Wrobel, 
N.~Xiang, 
W.~Ye.

\end{acknowledgments}

\appendix*

\section{Faster convergence with cross-sectional distribution}
\label{sec:betterConvergence}

The wakefield potential series can be made to converge faster 
(as $n$ increases, and/or as $m$ increases) by distributing the
driving charge in the radial and azimuthal directions, as 
well as the longitudinal direction.

To consider a distribution in $\theta$, one has to add the
beam angle $\theta_b$ to the wake potential; this is done merely
by replacing $\theta_t \rightarrow \theta_t-\theta_b$.
One can then convolute the bunch potential with
a normalized Gaussian distribution in $\theta_b$ to get an 
analytical result, as long
as the distribution is narrow enough that the tails that extend beyond 
$(\theta_b-\pi, \theta_b+\pi)$ are negligible.

To get an analytical result for a radial distribution is more difficult,
but we have found a distribution that allows fairly convenient computation.
We consider the function
\begin{equation}
	G(r;r_0,\sigma_r) 
	:= \frac{\displaystyle \exp\left[ -\frac{r_0(1+r_0/\sigma_r)}{r} - r/\sigma_r
	               \right] }{
	  2 r K_0 \left( 2\sqrt{(1+r_0/\sigma_r)r_0/\sigma_r} \right)
	        }
	= \frac{\displaystyle \exp\left[ 
	   -3\frac{r_0}{r} - \frac{(r-r_0)^2}{\sigma_r r}
	               \right] }{
	  2 r K_0 \left( 2\sqrt{(1+r_0/\sigma_r)r_0/\sigma_r} \right)
	        }
\end{equation}
($K_m$ is a modified Bessel function of order $m$)
with the following properties (for integrals, see
Ref.~\cite{Gradshteyn}, 3.471.9 and 6.635.3):
\begin{eqnarray}
      G(r_0) &=& 
	 \frac{ \exp\left[ -1 -2 r_0/\sigma_r)  
	               \right] }{
	  2 r_0 K_0 \left( 2\sqrt{(1+r_0/\sigma_r)r_0/\sigma_r} \right)
	        } \\
	G'(r_0) &=& 0 \\
	G''(r_0) &=& - \frac{1 + 2 r_0/\sigma_r}{r_0^2} G(r_0) \\
	\int_0^\infty dr \, G(r) &=& 1 \\
 \label{eq:integratedJm}
	IG(k) &\equiv& \int_0^\infty dr \, G(r) J_m(k r) \\
	    &=& \textstyle
	J_m \left(
	   \sqrt{2\frac{r_0}{\sigma_r} \left(1+\frac{r_0}{\sigma_r} \right)
	     \left(
	       \sqrt{\sigma_r^2 k^2+1} - 1 
	     \right)}
	    \right)
	    \displaystyle
	\frac{  K_m \left(
	   \sqrt{2\frac{r_0}{\sigma_r} \left(1+\frac{r_0}{\sigma_r} \right)
	     \left(
	       \sqrt{\sigma_r^2 k^2+1} + 1 
	     \right)}
	    \right)
	     }{
	  K_0 \left(
	   2\sqrt{(1+r_0/\sigma_r)r_0/\sigma_r}
	    \right) }
	    \nonumber \\
	IG(k) & \sim &
	J_m \left( k r_0 \right)  \qquad 
	 \textrm{ when } k \ll 1/\sigma_r \textrm{ and } r_0 \gg \sigma_r \\
	IG(k) & \sim &
	J_m \left( \sqrt{2k/\sigma_r} r_0 \right)
	\frac{
	   K_m \left( \sqrt{2k/\sigma_r} r_0 \right)
	     }{
	   K_0 \left( 2r_0/\sigma_r \right)
	    }  \nonumber \\
	  &\sim &
	\sqrt{ \frac{2}{\pi kr_0} }
	\exp \left[\left(2/\sigma_r - \sqrt{ 2k/\sigma_r }\right)r_0 \right]
	\\
	 && \textrm{ when } k \gg 1/\sigma_r \textrm{ and } r_0 \gg \sigma_r 
	 \nonumber
.\end{eqnarray}
That is, $G(r;r_0,\sigma_r)$ peaks at $r=r_0$, has unity integral,
approaches $\delta(r-r_0)$ for $\sigma_r\rightarrow 0$, 
and (most important) yields a not-too-complicated analytical result when
integrated with $J_m(kr)$.

To find the bunch potential for a radial distribution given by
$G(r;r_0,\sigma_r)$ one merely needs to replace $J_m(j_{m,n}r_b/R)$
in Eq.~(\ref{eq:bunchPotential}) with Eq.~(\ref{eq:integratedJm}),
where $r=r_b$ and $k=j_{m,n}/R$.

There are a couple of numerical difficulties to consider.  First,
one should use the identity
\begin{equation}
	       \sqrt{
	  \frac{j_{m,n}^2\sigma_r^2}{R^2}+1} - 1 
	     = \frac{j_{m,n}^2 \sigma_r^2/R^2}{
 	  \displaystyle     \sqrt{
	  \frac{j_{m,n}^2\sigma_r^2}{R^2}+1} \: + \: 1 
	     }
\end{equation}
in case $j_{m,n}\sigma_r/R \gg 1$.
Second, $K_m(x)$ and $K_0(y)$ become small enough to underflow
numerical arithmetic as $x$ and
$y$ become large, although their quotient may be of tractable magnitude.
In this case one had better evaluate:
\begin{equation}
	\frac{K_m(x)}{K_0(y)} = e^{y-x} \frac{e^x K_m(x)}{e^y K_0(y)}
\end{equation}
where the asymptotic expansion (see \cite{Abramowitz}) is used
to evaluate 
\begin{equation}
	e^x K_\nu ( x) = \sqrt{\frac{\pi}{2x}} \left[
	1 +
	\sum_{j=1}^\infty \frac{(4\nu^2 - 1)(4\nu^2-3^2)
	                        \cdots(4\nu^2-(2j-1)^2)}{
	  (8z)^j} \right]
\end{equation}
for $\nu=0,1$ (summing up to $j=5$ yields nearly full accuracy
for double precision when $x>400$), 
and then recursion is used for higher $\nu$:
\begin{equation}
	e^x K_{\nu+1}(x) = e^x K_{\nu-1}(x) - \frac{2\nu}{x} e^x K_{\nu}(x)
.\end{equation}


\begin{thebibliography}{16}
\expandafter\ifx\csname natexlab\endcsname\relax\def\natexlab#1{#1}\fi
\expandafter\ifx\csname bibnamefont\endcsname\relax
  \def\bibnamefont#1{#1}\fi
\expandafter\ifx\csname bibfnamefont\endcsname\relax
  \def\bibfnamefont#1{#1}\fi
\expandafter\ifx\csname citenamefont\endcsname\relax
  \def\citenamefont#1{#1}\fi
\expandafter\ifx\csname url\endcsname\relax
  \def\url#1{\texttt{#1}}\fi
\expandafter\ifx\csname urlprefix\endcsname\relax\def\urlprefix{URL }\fi
\providecommand{\bibinfo}[2]{#2}
\providecommand{\eprint}[2][]{\url{#2}}

\bibitem[{\citenamefont{Bane et~al.}(1984)\citenamefont{Bane, Wilson, and
  Weiland}}]{BaneWakeFieldGuide}
\bibinfo{author}{\bibfnamefont{K.~L.~F.} \bibnamefont{Bane}},
  \bibinfo{author}{\bibfnamefont{P.~B.} \bibnamefont{Wilson}},
  \bibnamefont{and} \bibinfo{author}{\bibfnamefont{T.}~\bibnamefont{Weiland}},
  \bibinfo{type}{Tech. Rep.} \bibinfo{number}{SLAC-PUB-3528},
  \bibinfo{institution}{SLAC} (\bibinfo{year}{1984}).

\bibitem[{\citenamefont{Wilson}(1989)}]{WilsonWakeFieldGuide}
\bibinfo{author}{\bibfnamefont{P.~B.} \bibnamefont{Wilson}},
  \bibinfo{type}{Tech. Rep.} \bibinfo{number}{SLAC-PUB-4547},
  \bibinfo{institution}{SLAC} (\bibinfo{year}{1989}).

\bibitem[{\citenamefont{Weiland and Zotter}(1981)}]{Weiland:1981}
\bibinfo{author}{\bibfnamefont{T.}~\bibnamefont{Weiland}} \bibnamefont{and}
  \bibinfo{author}{\bibfnamefont{B.}~\bibnamefont{Zotter}},
  \bibinfo{journal}{Part. Accel.} \textbf{\bibinfo{volume}{11}},
  \bibinfo{pages}{143} (\bibinfo{year}{1981}).

\bibitem[{\citenamefont{Ratschow and Weiland}(2002)}]{Ratschow:2002}
\bibinfo{author}{\bibfnamefont{S.}~\bibnamefont{Ratschow}} \bibnamefont{and}
  \bibinfo{author}{\bibfnamefont{T.}~\bibnamefont{Weiland}},
  \bibinfo{journal}{Phys. rev. spec. top., Accel. beams}
  \textbf{\bibinfo{volume}{5}}, \bibinfo{pages}{052001} (\bibinfo{year}{2002}).

\bibitem[{\citenamefont{Tsakanian}(2005)}]{Tsakanian:2005}
\bibinfo{author}{\bibfnamefont{A.}~\bibnamefont{Tsakanian}},
  \bibinfo{journal}{Nucl. Instrum. Methods A.} \textbf{\bibinfo{volume}{548}},
  \bibinfo{pages}{298} (\bibinfo{year}{2005}).

\bibitem[{\citenamefont{Kim et~al.}(1990)\citenamefont{Kim, Chen, and
  Yang}}]{Kim:1990}
\bibinfo{author}{\bibfnamefont{S.~H.} \bibnamefont{Kim}},
  \bibinfo{author}{\bibfnamefont{K.~W.} \bibnamefont{Chen}}, \bibnamefont{and}
  \bibinfo{author}{\bibfnamefont{J.~S.} \bibnamefont{Yang}},
  \bibinfo{journal}{J. Appl. Phys.} \textbf{\bibinfo{volume}{68}},
  \bibinfo{pages}{4942} (\bibinfo{year}{1990}).

\bibitem[{\citenamefont{Gao}(1995)}]{Gao:1995}
\bibinfo{author}{\bibfnamefont{J.}~\bibnamefont{Gao}}, in
  \emph{\bibinfo{booktitle}{1995 Particle Accelerator Conf.}}
  (\bibinfo{year}{1995}), p. \bibinfo{pages}{1055}.

\bibitem[{\citenamefont{Gao}(1996)}]{Gao:1996}
\bibinfo{author}{\bibfnamefont{J.}~\bibnamefont{Gao}}, \bibinfo{journal}{Nucl.
  Instrum. Methods A.} \textbf{\bibinfo{volume}{381}}, \bibinfo{pages}{174}
  (\bibinfo{year}{1996}).

\bibitem[{\citenamefont{Weiland}(1983)}]{Weiland:1983}
\bibinfo{author}{\bibfnamefont{T.}~\bibnamefont{Weiland}},
  \bibinfo{journal}{Nucl. Instrum. Methods} \textbf{\bibinfo{volume}{216}},
  \bibinfo{pages}{31} (\bibinfo{year}{1983}).

\bibitem[{\citenamefont{Napoly et~al.}(1993)\citenamefont{Napoly, Chin, and
  Zotter}}]{Napoly:1993}
\bibinfo{author}{\bibfnamefont{O.}~\bibnamefont{Napoly}},
  \bibinfo{author}{\bibfnamefont{Y.~H.} \bibnamefont{Chin}}, \bibnamefont{and}
  \bibinfo{author}{\bibfnamefont{B.}~\bibnamefont{Zotter}},
  \bibinfo{journal}{Nucl. Instrum. Methods A.} \textbf{\bibinfo{volume}{334}},
  \bibinfo{pages}{255} (\bibinfo{year}{1993}).

\bibitem[{\citenamefont{Chao and Tigner}(1999)}]{AcceleratorHandbook}
\bibinfo{editor}{\bibfnamefont{A.~W.} \bibnamefont{Chao}} \bibnamefont{and}
  \bibinfo{editor}{\bibfnamefont{M.}~\bibnamefont{Tigner}}, eds.,
  \emph{\bibinfo{title}{Handbook of Accelerator Physics and Engineering}}
  (\bibinfo{publisher}{World Scientific}, \bibinfo{year}{1999}).

\bibitem[{\citenamefont{Abramowitz and Stegun}(1972)}]{Abramowitz}
\bibinfo{editor}{\bibfnamefont{M.}~\bibnamefont{Abramowitz}} \bibnamefont{and}
  \bibinfo{editor}{\bibfnamefont{I.~A.} \bibnamefont{Stegun}}, eds.,
  \emph{\bibinfo{title}{Handbook of Mathematical Functions}}
  (\bibinfo{publisher}{Dover Publications}, \bibinfo{year}{1972}).

\bibitem[{\citenamefont{Weideman}(1994)}]{Weideman:1994}
\bibinfo{author}{\bibfnamefont{J.~A.~C.} \bibnamefont{Weideman}},
  \bibinfo{journal}{SIAM J. Numer. Anal.} \textbf{\bibinfo{volume}{31}},
  \bibinfo{pages}{1497} (\bibinfo{year}{1994}).

\bibitem[{\citenamefont{Nieter and Cary}(2004)}]{Nieter:2004}
\bibinfo{author}{\bibfnamefont{C.}~\bibnamefont{Nieter}} \bibnamefont{and}
  \bibinfo{author}{\bibfnamefont{J.~R.} \bibnamefont{Cary}},
  \bibinfo{journal}{J. Comput. Phys.} \textbf{\bibinfo{volume}{196}},
  \bibinfo{pages}{448} (\bibinfo{year}{2004}).

\bibitem[{\citenamefont{Dey and Mittra}(1999)}]{DeyMittra:1999}
\bibinfo{author}{\bibfnamefont{S.}~\bibnamefont{Dey}} \bibnamefont{and}
  \bibinfo{author}{\bibfnamefont{R.}~\bibnamefont{Mittra}},
  \bibinfo{journal}{IEEE Trans. Microwave Theory Tech.}
  \textbf{\bibinfo{volume}{47}}, \bibinfo{pages}{1737} (\bibinfo{year}{1999}).

\bibitem[{\citenamefont{Gradshteyn and Ryzhik}(1994)}]{Gradshteyn}
\bibinfo{author}{\bibfnamefont{I.~S.} \bibnamefont{Gradshteyn}}
  \bibnamefont{and} \bibinfo{author}{\bibfnamefont{I.~M.}
  \bibnamefont{Ryzhik}}, \emph{\bibinfo{title}{Tables of Integrals, Series, and
  Products}} (\bibinfo{publisher}{Academic Press, Inc.}, \bibinfo{year}{1994}),
  \bibinfo{edition}{5th} ed.

\end{thebibliography}

\end{document}